\documentclass[journal,twocolumn,10pt]{IEEEtran}
\usepackage{graphicx}
\usepackage{amssymb}
\usepackage{mathtools}
\usepackage{dsfont}
\usepackage{cite}
\usepackage{stfloats}
\usepackage{subfigure}
\usepackage{psfrag}
\usepackage[mathscr]{euscript}
\usepackage{acronym}  
\usepackage{amsmath}
\usepackage{algorithm}
\usepackage[noend]{algpseudocode}
\usepackage{booktabs}
\usepackage{bbm}

\makeatletter
\def\BState{\State\hskip-\ALG@thistlm}
\makeatother

\usepackage{float}
\usepackage{balance}

\acrodef{CCDF}{complementary cumulative distribution function}
\acrodef{CF}{characteristic function}
\acrodef{PPP}{Poisson point processe}
\acrodef{RV}{random variable}
\acrodef{i.i.d.}{independent and identically distributed}
\acrodef{PDF}{probability distribution function}
\acrodef{CDF}{cumulative distribution function}
\acrodef{ch.f.}{characteristic function}
\acrodef{AWGN}{additive white Gaussian noise}
\acrodef{SNR}{signal-to-noise ratio}
\acrodef{LRT}{likelihood ratio test}
\acrodef{DRT}{distance ratio test}
\acrodef{GLRT}{generalized likelihood ratio test}
\acrodef{CRLB}{Cram\'{e}r-Rao lower bound}
\acrodef{CRB}{Cram\'{e}r-Rao bound}
\acrodef{ZZLB}{Ziv-Zakai lower bound}
\acrodef{ZZB}{Ziv-Zakai bound}
\acrodef{LOS}{line-of-sight}
\acrodef{ToF}{time-of-flight}
\acrodef{NLOS}{non-line-of-sight}
\acrodef{GDOP}{geometric dilution of precision}
\acrodef{GPS}{Global Positioning System}
\acrodef{FIM}{Fisher information matrix}
\acrodef{PEB}{position error bound}
\acrodef{SPEB}{squared position error bound}
\acrodef{TOA}{time-of-arrival}
\acrodef{TOF}{time-of-flight}
\acrodef{WSN}{wireless sensor network}
\acrodef{MAC}{medium access control}
\acrodef{RSS}{received signal strength}
\acrodef{WAF}{wall attenuation factor}
\acrodef{TDOA}{time difference-of-arrival}
\acrodef{RF}{radiofrequency}
\acrodef{RTT}{round-trip time}
\acrodef{AOA}{angle-of-arrival}
\acrodef{MF}{matched filter}
\acrodef{ED}{energy detector}
\acrodef{ML}{maximum likelihood}
\acrodef{MSE}{mean-square error}
\acrodef{RMSE}{root-mean-square error}
\acrodef{LEO}{localization error outage}
\acrodef{ppm}{part-per-million}
\acrodef{ACK}{acknowledge}
\acrodef{UWB}{Ultrawide bandwidth}
\acrodef{TNR}{threshold-to-noise ratio}
\acrodef{LS}{least squares}
\acrodef{IR-UWB}{impulse radio UWB}
\acrodef{FCC}{Federal Communications Commission}
\acrodef{TH}{time-hopping}
\acrodef{PPM}{pulse position modulation}
\acrodef{MUI}{multi-user interference}
\acrodef{PDP}{power delay profile}
\acrodef{BPZF}{band-pass zonal filter}
\acrodef{SIR}{signal-to-interference ratio}
\acrodef{SINR}{signal-to-interference-plus-noise ratio}
\acrodef{RFID}{radio frequency identification}
\acrodef{WPAN}{wireless personal area network}
\acrodef{WWB}{Weiss-Weinstein bound}
\acrodef{DP}{direct path}
\acrodef{MF}{matched filter}
\acrodef{MMSE}{minimum-mean-square-error}
\acrodef{SBS}{serial backward search}
\acrodef{SBSMC}{serial backward search for multiple clusters}
\acrodef{NBI}{narrowband interference}
\acrodef{WBI}{wideband interference}
\acrodef{INR}{interference-to-noise ratio}
\acrodef{CR}{channel response}
\acrodef{CIR}{channel impulse response}
\acrodef{CR}{channel  response}
\acrodef{RADAR}{radar}
\acrodef{MUR}{Multistatic radar}
\acrodef{JBSF}{jump back and search forward}
\acrodef{HDSA}{high-definition situation-aware}
\acrodef{RRC}{root raised cosine}
\acrodef{ST}{simple thresholding}
\acrodef{BTB}{Bellini-Tartara bound}
\acrodef{P-Max}{$P$-Max}  
\acrodef{MIMO}{multiple-input multiple-output}
\acrodef{MAP}{maximum a posteriori}
\acrodef{FG}{factor graph}
\acrodef{OP}{outage probability}
\acrodef{WED}{wall extra delay}
\acrodef{RMS}{root mean square}
\acrodef{SPAWN}{sum-product algorithm over a wireless network}
\acrodef{MDD}{minimum distance distribution}
\acrodef{MAP}{maximum a posteriori probability}
\acrodef{SAP}{small cell access point}
\acrodef{UE}{user equipment}
\acrodef{MBS}{macro cell base station}
\acrodef{UER}{\ac{UE} Relay}
\acrodef{D2D}{device-to-device}
\acrodef{MBS}{macro base station}
\acrodef{CSI}{channel state information}
\acrodef{OGR}{outage guard region}
\acrodef{FUR}{feasible UER region}
\acrodef{EHR}{energy harvesting region}
\acrodef{EH}{energy harvesting}
\acrodef{D2D-EHSN}{D2D communication provided \ac{EH} small cell network}
\acrodef{D2D-EHHN}{D2D communication provided \ac{EH} heterogeneous network}
\acrodef{3GPP}{3rd Generation Partnership Project}
\acrodef{BS}{base station}
\acrodef{DF}{decode and forward}
\acrodef{CCDF}{complementary cumulative distribution function}
\acrodef{ZF}{zero forcing}
\acrodef{RZF}{regularized zero forcing}
\acrodef{WLLN}{weak law of large number}
\acrodef{SLLN}{strong law of large numbers}
\acrodef{TDD}{Time-division duplex}
\acrodef{EE}{energy efficiency} 
\acrodef{HetNet}{heterogeneous network} 
\acrodef{SCP}{Single Cell Processing}
\acrodef{CBF}{Coordinated Beamforming}
\usepackage{color}
\usepackage{dsfont}
\usepackage{bbm}

\DeclareMathAlphabet{\mathsf}{OML}{cmbr}{m}{it}

\newtheorem{definition}{\bf Definition}
\newtheorem{theorem}{\bf Theorem}
\newtheorem{lemma}{\bf Lemma}

\newtheorem{remark}{\bf Remark}

\newtheorem{assumption}{\bf Assumption}

\newcommand{\bd}{\begin{description}}
\newcommand{\ed}{\end{description}}
\newcommand{\be}{\begin{enumerate}}
\newcommand{\ee}{\end{enumerate}}
\newcommand{\bi}{\begin{itemize}}
\newcommand{\ei}{\end{itemize}}
\newcommand{\bl}{\begin{list}}
\newcommand{\el}{\end{list}}
\newcommand{\bt}{\begin{tabbing}}
\newcommand{\et}{\end{tabbing}}

\newcommand{\paperTitle}{ Age of Information in Random Access Networks: A Spatiotemporal Study }

\begin{document}

{
\title{\paperTitle}

\author{

	    Howard~H.~Yang$^\mathsection$$^\dagger$,
        Ahmed Arafa$^\ast$,
	    Tony~Q.~S.~Quek$^\dagger$,
	    and H.~Vincent~Poor$\ddagger$\\
       $^\mathsection$ \textit{Zhejiang University/University of Illinois at Urbana-Champaign Institute, Haining 314400, China }\\
       $^\dagger$ \textit{Singapore University of Technology and Design, Singapore 487372}\\
       $^\ast$ \textit{University of North Carolina at Charlotte, NC 28223, USA} \\
       $\ddagger$ \textit{Princeton University, Princeton, NJ 08544, USA}
\thanks{This work was supported in part by the U.S. National Science Foundation under Grant CCF-1908308.}


%
%
\vspace{-.2in}
}
\maketitle
\acresetall
\thispagestyle{empty}
\begin{abstract}
We investigate the age-of-information (AoI) in the context of random access networks, in which transmitters need to send a sequence of information packets to intended receivers over shared spectrum. We establish an analytical framework that accounts for the key features of a wireless system, including the fading, path loss, network topology, as well as the spatial interactions amongst the queues. A closed-form expression is derived to quantify the network average AoI and its accuracy is verified via simulations. Our analysis unveils several unconventional behaviors of AoI in such a setting. For instance, even when the packet transmissions are scheduled in a last-come first-serve (LCFS) order whereby the newly incoming packets can replace the undelivered ones, the network average AoI may not monotonically decline with respect to the packet arrival rates, if the infrastructure is densely deployed. Moreover, the ALOHA protocol is shown to be instrumental in reducing the AoI when the packet arrival rates are high, yet it cannot contribute to decreasing the AoI in the regime of infrequent packet arrivals.
\end{abstract}
%

\acresetall

\section{Introduction}\label{sec:intro}
Fueled by the eagerness for fresh data in many real-time applications, the age-of-information (AoI) has been introduced as a metric to assess the ``freshness'' of information delivered over a period of time \cite{KauYatGru:12}.
Compared with transmitter-centric metrics, e.g., delay or throughput, AoI puts the focus on the receiver side and measures the time elapsed since the latest packet has been delivered, thus being able to gauge the “freshness” associated with the information packets \cite{KosPapAng:17,SunUysYat:17,YatKau:18,ZhaAraHua:19,AraYanUlu:18,WuYanWu:18,BacSunUys:19}. As such, networks designed by minimizing the metric of AoI enable the acquisition of fresh data and are particularly relevant in Internet of Things (IoT) applications where the timeliness of information is crucial, e.g., monitoring the status of a system or asserting remote controls based on information collected from a network of sensors \cite{AbdPapDhi:19,ZhoSaa:18,XuYanWan:20,XuWanYan:20INFOCOM}.
Because these platforms generally constitute a random access network in which multiple source nodes need to communicate with their destinations via shared spectrum, the interference amongst transmitters located in geographical proximity may be severe and lead to transmission failures that hinder the timely updates of information.
In response, a number of strategies to schedule the set of simultaneously active links have been proposed \cite{HeYuaEph:16,talak2018optimizing}, that achieve age minimization by limiting the interference to an acceptable range.
Moreover, several threshold-based channel access schemes have also been developed to optimize the AoI from a network perspective \cite{CheGatHas:19,CheGuLie:20}.
However, these results are devised based on collision models or conflict graphs, which do not precisely capture the key attributes of a wireless system such as fading, path loss, and co-channel interference.
Recognizing this, a recent line of research has been carried out \cite{HuZhoZha:18,EmaElSBau:20,YanArafaQue:19}, that conflates queueing theory with stochastic geometry -- a disruptive tool for assessing the performance of wireless links in large-scale networks -- to account for the spatial, temporal, and physical level attributes in the analysis of AoI. Consequently, lower and upper bounds on the distribution of average AoI are derived in the context of a Poisson network \cite{HuZhoZha:18}. Additionally, the performance of peak AoI in uplink IoT networks is analytically evaluated under time-triggered and event-triggered traffic profiles \cite{EmaElSBau:20}. Moreover, a distributed algorithm that configures the channel access probabilities at each individual transmitter based on the local observation of the network topology is proposed to minimize the peak AoI \cite{YanArafaQue:19}.
Nonetheless, these works assume that transmissions of information packets are scheduled in a first-come first-serve (FCFS) discipline which is not appealing for minimizing the AoI. Moreover, the performance metric considered in most of these works is the \textit{peak} AoI while the more commonly used metric of \textit{average} AoI has not been well-studied.

In this paper, we aim to develop a theoretical template for a thorough understanding of the AoI over a random access network. Toward that purpose, we model the positions of transmitter-receiver pairs as a Poisson bipolar network. Each transmitter generates a sequence of status updates, encapsulated in the information packets, according to independent Bernoulli processes.
The newly incoming packets at each transmitter are stored in a unit-size buffer and replace the older undelivered ones, if any.
In each time slot, transmitters with non-empty buffers employ an ALOHA protocol to access the shared spectrum and send out packets when granted approval.
Different from most related works, the transmissions are successful only if the received signal-to-interference-plus-noise ratio (SINR) exceeds a decoding threshold, upon which the packet can be removed from the transmitter buffer.
Otherwise, the packet stays in the buffer and will be retransmitted in the next available time slot (unless replaced by a new generated packet).
Because of interference, there is a coupling between the node position and its active state. We thus jointly use tools from stochastic geometry, to capture the macroscopic interference behavior, and queueing theory, to characterize the evolution of queues at the microscopic level, to derive accurate and closed-form expressions for the network average AoI.
The analytical results enable us to explore the effects of different network parameters on the AoI performance and hence can serve as useful guidelines for further system designs.

\section{System Model}\label{sec:sysmod}

\subsection{Spatial Configuration and Physical Layer Parameters }
Let us consider a wireless network that consists of a set of transmitter-receiver pairs, all located in the Euclidean plane.
The transmitters are scattered according to a homogeneous Poisson point process (PPP) $\tilde{\Phi}$ of spatial density $\lambda$, where a generic node $i$ located at $X_i \in \tilde{\Phi}$ has one dedicated receiver at $y_i$, which is at distance $r$ to $X_i$ and oriented in a uniformly random direction\footnote{Such a setting is commonly known as the Poisson bipolar model \cite{BacBla:09}, which is a large-scale analog to the classical model of Random Networks \cite{GupKum:00} and has been widely used for the modeling of networks without a centralized infrastructure, e.g., the D2D, IoT, and wireless ad-hoc networks.}.
According to the displacement theorem \cite{BacBla:09}, the location set $\bar{\Phi} = \{y_i\}_{i=0}^\infty$ also forms a homogeneous PPP with spatial density $\lambda$.
If a transmitter needs to communicate with its receiver, it employs a unified power $P_{\mathrm{tx}}$ and sends out packets over a shared spectrum, which is affected by Rayleigh fading with unitary power gain and path-loss that follows power law attenuation.
All channel gains are independent and identically distributed (i.i.d.) across space and time.
Besides, the transmission is also subject to Gaussian thermal noise with variance $\sigma^2$.

\subsection{Temporal Configuration and Transmission Protocol }
We assume the network is synchronized\footnote{Synchronization over networks can be achieved by either centralized \cite{WanLinAdh:17} or distributed mechanisms \cite{XioWuShe:17}. } and the time is segmented into slots with each equal to the duration to finish one packet transmission. At the beginning of each time slot, every transmitter has an arrival of information packet with probability $\xi \in (0,1]$.
The newly incoming packet at each transmitter will be stored in a unit-size buffer and replace the undelivered older one if there is any.
In that respect, the schedule of packet transmissions constitutes a last-come first-serve with replacement (LCFS-R) protocol.

In each time slot, transmitters with non-empty buffers adopt the ALOHA protocol with probability $p$ to access the radio channel, and send out packets when granted approval.
A transmission is considered successful if the SINR received at the destination exceeds a decoding threshold, upon which the receiver sends an ACK feedback message so that the packet can be removed from the buffer.
Otherwise, the receiver sends a NACK feedback message and the packet is retransmitted in the next available time slot\footnote{We assume the ACK/NACK transmission is instantaneous and error-free, as commonly done in the literature \cite{TalKarMod:18}.}.
In this network, the delivery of packets incurs a delay of one time slot, namely, packets are transmitted at the beginning of time slots and, if the transmission is successful, they are delivered by the end of the same time slot.

Because the time scale of fading and packet transmission is much smaller than that of the spatial dynamics, we assume the network topology is static, i.e., an arbitrary but fixed point pattern is realized at the beginning and remains unchanged over the time domain.

\begin{figure}[t!]
  \centering{}

    {\includegraphics[width=0.95\columnwidth]{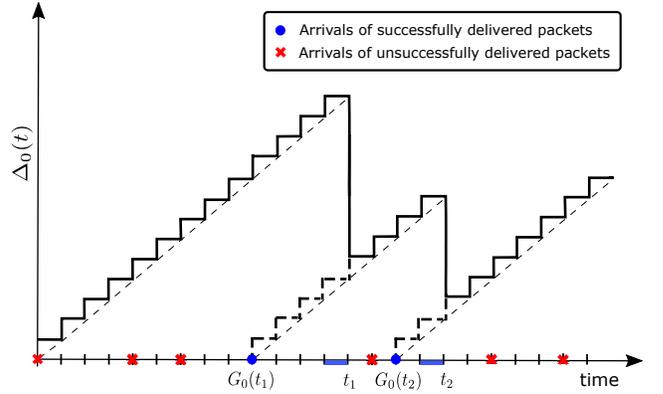}}

  \caption{ AoI evolution example at a typical link under the LCFS-R discipline. The time instances $G_0(t_i)$ and $t_i$ respectively denote the moments when the $i$-th packet is generated and delivered, and the age is reset to $t_i - G_0(t_i)$. }
  \label{fig:AoIMod_V1}
\end{figure}
\subsection{Age of Information}
The performance metric of this work is the AoI, which captures the timeliness of information delivered at the receiver side.
A formal definition of this notation is stated in below.
\begin{definition}
\textit{Consider a typical transmitter-receiver pair. Let $\{G(t_i)\}_{i \geq 1}$ be the sequence of generation times of information packets and $\{t_i\}_{ i \geq 1 }$ be the corresponding times at which these packets are received at the destination. Amongst the packets received till time $t$, denote the index of the latest generated one by $n_t = \arg\max_{i}\{ G(t_i) | t_i \leq t \}$. The age of information at the receiver is defined as $\Delta(t) = t - G(t_{n_t})$.
}
\end{definition}

If the average time for packet delivery is the same, then according to Definition~1, the presence of a new packet at the transmitter will make the older one irrelevant in reducing the AoI.
As such, maintaining a unit-size buffer at each transmitter and replacing undelivered packets with newly incoming ones is consistent with the minimum AoI packet management strategy.

Without loss of generality, we denote the link pair located at $(X_0, y_0)$, where $y_0$ is the origin, as typical.
Under the employed system model, the AoI over the typical link goes up by one in each time slot if no new packet is updated at the receiver side, and, when the update is received, reduces to the time elapsed since the generation of the delivered packet. An example of the dynamics of AoI is illustrated in Figure~\ref{fig:AoIMod_V1}.
Formally, the evolution of $\Delta_0(t)$ can be written as follows:
\begin{align*}
\Delta_0(t \!+\! 1) =
\left\{
       \begin{array}{ll}
         \!\!  \Delta_0(t)  +   1, \quad \quad ~~   \text{if no update received}, \\
         \!\!  t  -  G_0(t) + 1, \quad \quad ~~ \text{otherwise}
       \end{array}
\right.
\end{align*}
where $G_0(t)$ is the generation time of the packet delivered over the typical link by the end of time slot $t$.

In this work, we use the \textit{average} AoI as our metric to evaluate the freshness of information over the considered random access network. Specifically, the average AoI at a given link $j$ is defined as
\begin{align}
\bar{\Delta}_j = \limsup_{ T \rightarrow \infty } \frac{ 1 }{ T } \sum_{ t=1 }^T \Delta_j(t).
\end{align}
By extending this concept to a large scale, we define the \textit{network} average AoI as follows:
\begin{align}
\bar{ \Delta } &= \limsup_{R \rightarrow \infty} \frac{ \sum_{ X_j \in \tilde{\Phi} \cap B(0,R) } \bar{\Delta}_j }{ \sum_{ X_j \in \tilde{\Phi} } \mathbbm{1}\{ X_j \!\in\! B(0,R) \}  }
\nonumber\\
&\stackrel{(a)}{=}   \mathbb{E}^0 \Big[ \limsup_{ T \rightarrow \infty } \frac{1}{T} \sum_{t=1}^T \Delta_0( t )  \Big]
\end{align}
where $B(0,R)$ denotes a disk centered at the origin with radius $R$, $\mathbbm{1\{ \cdot \}}$ is the indicator function, and $(a)$ follows from Campbell's theorem \cite{BacBla:09}. The notion $\mathbb{E}^0[\cdot]$ indicates that the expectation is taken with respect to the Palm distribution $\mathbb{P}^0$ of the stationary point process -- the condition will be given in Section~III -- where under $\mathbb{P}^0$ almost surely there is a node located at the origin \cite{BacBla:09}.

\section{ Analysis }
This section constitutes the main technical part of our paper, in which we derive analytical expressions to characterize the statistics of AoI.
For better readability, most proofs and mathematical derivations have been relegated to the Appendix.
\subsection{Preliminaries}
\subsubsection{SINR at a typical receiver}
Due to the stationary property of PPPs, we can apply Slivnyak's theorem \cite{BacBla:09} and concentrate on a \textit{typical receiver} located at the origin, with its tagged transmitter situated at $X_0$. Note that when averaging over the point process, this representative link has the same statistic as those obtained by averaging over other links in the network.
As such, if the transmitter sends out a packet during time slot $t$, the SINR received at the destination can be written as
\begin{align} \label{equ:SINR_expression}
\mathrm{SINR}_{0,t} = \frac{P_{\mathrm{tx}} H_{00} r^{-\alpha} }{ \sum_{ j \neq 0 } P_{\mathrm{tx}} H_{j0} \zeta_{j,t} \Vert X_j \Vert^{-\alpha} + \sigma^2 }
\end{align}
where $\alpha$ denotes the path loss exponent, $H_{ji} \sim \exp(1)$ is the channel fading coefficient from transmitter $j$ to receiver $i$ which varies in each time slot, $\zeta_{j,t} \in \{ 0, 1 \}$ is an indicator showing whether node $j$ is active ($\zeta_{j,t}=0$) or not ($\zeta_{j,t}=1$).
\subsubsection{Conditional transmission success probability}
Seen from the temporal perspective, dynamics on any given link can be abstracted as a Geo/G/1/2 queue with replacement in which the service rate is dependent on the statistics of SINR, namely the transmission success probability.
Because the network is considered to be static, we condition on the node positions $\Phi \triangleq \tilde{\Phi} \cup \bar{\Phi}$ and define the conditional transmission success probability of the typical link at time slot $t$ as follows \cite{YanQue:19}
\begin{align}\label{equ:CndTX_Prob}
\mu^\Phi_{0,t} = \mathbb{P}\big(\mathrm{SINR}_{0,t} > \theta | \Phi\big)
\end{align}
where $\theta$ is the decoding threshold.

Due to the broadcast nature of wireless medium, transmissions over the link pairs are correlated in both space and time via the interference they cause.
This phenomenon is usually referred to as the spatially interacting queues \cite{SanBac:17}, which results in $\{\mu^\Phi_{0,t}\}_{t \geq 0}$ being correlated over time and hinders tractable analysis.
For the sake of tractability, we need the following approximation.
\begin{assumption}
\textit{Each node experiences independent interferers over time, and hence their queues evolve independently from each other.}
\end{assumption}
This assumption is commonly known as the \textit{mean-field approximation}, which has been shown to be applicable to the spatiotemporal analysis of large-scale networks \cite{ChiElSCon:19}.

\subsubsection{Conditional Age of Information}
Following Assumption~1, when we condition on the network topology $\Phi$, the transmissions of packets over a typical link are i.i.d. over time with a success probability $\mu^\Phi_0 = \lim_{t \rightarrow \infty} \mu^\Phi_{0,t}$.
As such, we can treat the dynamics at the typical sender as a Geo/Geo/1/2 queue where the arrival and departure rates are given by $\xi$ and $p \mu^\Phi_{0}$, respectively. In consequence, we can leverage tools from queueing theory and arrive at a conditional form of the AoI.
\begin{lemma} \label{lma:Cndt_AoI}
\textit{
	Conditioned on the point process $\Phi$, the average AoI at the typical link is given as follows:
	\begin{align} \label{equ:Cnd_AoI_LCFS}
	\mathbb{E}^0\!\big[\, \bar{\Delta}_0 \vert \Phi \,\big] \!&= \frac{1}{\xi} \!+\! \frac{1}{ p \mu^\Phi_0 } - 1.
	\end{align}
}
\end{lemma}
\begin{IEEEproof}
See Appendix~\ref{apn:Cndt_AoI}.
\end{IEEEproof}

In view of Lemma~\ref{lma:Cndt_AoI}, we note that the core of analyzing the AoI lies at the characterization of the transmission success probability. In the following, we detail the procedure of deriving this quantity.

\subsection{ Transmission Success Probability }
Using Assumption~1, we can now assume that each node activates independently in the steady state, and hence compute the conditional transmission success probability as follows.

\begin{lemma} \label{lma:Cnd_SucProb}
\textit{Conditioned on the network topology $\Phi$, the probability of successful transmission over the typical link is given as:
\begin{align}
\mu^\Phi_0 = e^{-\frac{ \theta r^\alpha}{\rho} } \prod_{ j \neq 0 } \Big( 1 - \frac{ a^\Phi_j }{ 1 + \Vert X_j \Vert^\alpha / \theta r^\alpha } \Big)
\end{align}
where $\rho = P_{\mathrm{tx}}/\sigma^2$ and $a^\Phi_j = \lim_{ T \rightarrow \infty} \sum_{t=0}^{T} \zeta_{j,t}/T$ is the active probability of node $j$ in the steady state.
}
\end{lemma}
\begin{IEEEproof}
See \cite{YanArafaQue:19} for a detailed proof.
\end{IEEEproof}

We can now explicitly identify the randomness in the departure rate, which mainly arises from $i$) the random locations of interfering nodes, and $ii$) their corresponding active states. A conditional expression for the active state at each communication link can be obtained as follows.

\begin{lemma} \label{lma:Cndt_ActProb}
\textit{Conditioned on the network topology $\Phi$, the active probability of a generic node $j$ is given as:
\begin{align}
a^\Phi_j = \frac{ p \, \xi }{ \xi + ( 1 - \xi ) \, p \, \mu^\Phi_j  }.
\end{align}
}
\end{lemma}
\begin{IEEEproof}
See Appendix~\ref{apn:Cndt_ActProb}.
\end{IEEEproof}

\setcounter{equation}{\value{equation}}
\setcounter{equation}{8}
\begin{figure*}[t!]
\begin{align} \label{equ:Meta_Grl}
F(u) = \frac{1}{2} -\! \int_{0}^{\infty} \!\!\! \mathrm{Im}\bigg\{ u^{-j\omega} \exp\!\Big(\! - \frac{ j \omega \theta r^\alpha }{ \rho } -  \lambda \pi r^2 \theta^\delta \sum_{k=1}^{\infty} \binom{j \omega}{ k } \!\int_0^\infty \frac{ (-1)^{k+1} dv }{ (1+v^{ \frac{ \alpha }{ 2 } } )^k }  \int_{0}^1 \!\! \frac{ (p \xi)^k F(dt) }{ [\, \xi + (1-\xi) p t \,]^k } \Big) \bigg\} \frac{ d \omega }{ \pi \omega }
\end{align}
\setcounter{equation}{\value{equation}}{}
\setcounter{equation}{9}
\centering \rule[0pt]{18cm}{0.3pt}
\end{figure*}
\setcounter{equation}{9}

With these results in hand, we can now put the pieces together and derive the distribution of the conditional transmission success probability.

\begin{theorem} \label{thm:Meta_SINR}
  \textit{
  The cumulative distribution function (CDF) of the conditional transmission success probability is given by the fixed-point equation \eqref{equ:Meta_Grl} at the top of the next page, in which $j=\sqrt{-1}$ and $\mathrm{Im}\{ \cdot \}$ denotes the imaginary part of a complex quantity.
  }
\end{theorem}
\begin{IEEEproof}
See Appendix~\ref{apn:Meta_SINR}.
\end{IEEEproof}

Owing to the space-time coupling amongst the queues, the transmission success probability CDF \eqref{equ:Meta_Grl} is given in the form of a fixed-point functional equation. It is noteworthy that the right hand side of \eqref{equ:Meta_Grl} constitutes a contraction as a functional of $F(\cdot)$. As such, solution of \eqref{equ:Meta_Grl} can be obtained via successive approximations \cite{YanQue:19}, i.e., the Picard's method, which converges exponentially fast.

\subsection{Network Average AoI}
We are now ready to present the main results of this paper, i.e., the analytical expressions for the AoI.
\begin{theorem} \label{thm:AoI_LCFS}
\textit{
	The network average AoI is given as follows:
	\begin{align} \label{equ:EctForm_AoI_LCFS}
	\bar{ \Delta } = \frac{1}{\xi} + \!\! \int_{0}^{1} \!\! \frac{ F(dt) }{ p t } -1
	\end{align}
    where $F(\cdot)$ is given in \eqref{equ:Meta_Grl}.
}
\end{theorem}
\begin{IEEEproof}
By deconditioning \eqref{equ:Cnd_AoI_LCFS} according to the CDF of $\mu^\Phi_0$ per \eqref{equ:Meta_Grl}, the result follows.
\end{IEEEproof}

Notably, the AoI expression in Theorem~\ref{thm:AoI_LCFS} accounts for all the key features of a random access network, including the packet arrival rate, channel access probability, deployment density, and interference.
We will verify the accuracy of this analysis in Section~IV and obtain a number of design insights based on numerical results.
Before that, let us remark a special case as follows.

\remark{\textit{ When $\lambda \rightarrow 0$, i.e., the network is in the noise-limited regime, it can be shown that the network average AoI is given by
\begin{align}
\bar{\Delta} = \frac{1}{\xi} + \frac{\exp\big( \frac{\theta r^\alpha}{\rho} \big)}{p} - 1,
\end{align}
which monotonically decreases with the packet arrival rate $\xi$.
}}

This observation is in line with conclusions drawn from the conventional point-to-point settings, namely under the LCFS discipline, increasing the update frequency can always benefit the AoI performance.

\section{Simulation and Numerical Results}
In this section, we show simulation results that confirm the accuracy of our analytical framework, and based on the analysis we further investigate the AoI performance under different settings of network parameters.
During each simulation run, we realize the node positions over a 1 $\text{km}^2$ area according to a Poisson bipolar model with spatial density $\lambda$. The packet arrivals at each source node are generated as independent Bernoulli processes with rate $\xi$.
We average over 10,000 realizations and collect the statistic from every link to calculate the average AoI.
Unless differently specified, we use the following parameters: $\alpha = 3.8$, $\theta=0$~dB, $P_{\mathrm{tx}}=17$~dBm, and $\sigma^2 = -90$~dBm.

\begin{figure}[t!]
  \centering{}

    {\includegraphics[width=0.95\columnwidth]{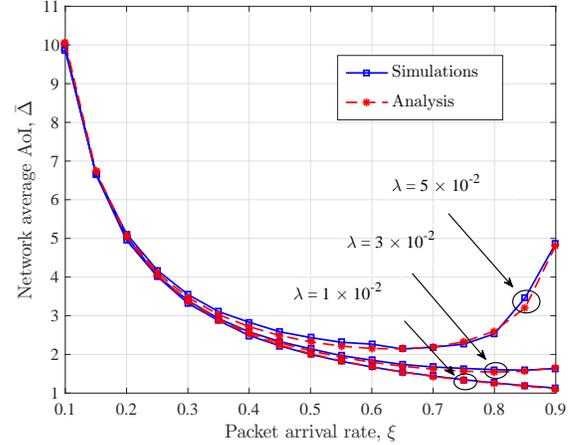}}

  \caption{ Simulation versus analysis of the network average AoI, in which we set $p=1$, $r = 0.5$ m, and vary the deployment densities as $\lambda = 1 \times 10^{-2}, 3 \times 10^{-2}, 5 \times 10^{-2}$ m$^{-2}$. }
  \label{fig:SimVerf}
\end{figure}

\begin{figure}[t!]
  \centering{}

    {\includegraphics[width=0.95\columnwidth]{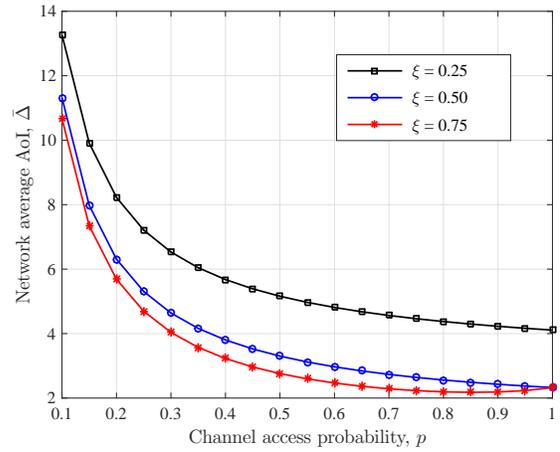}}

  \caption{ The average AoI versus channel access probability, where we set $r=0.5$ m, $\lambda = 5 \times 10^{-2}$ m$^{-2}$, and vary the packet update frequencies as $\xi = 0.25, 0.50, 0.75$. }
  \label{fig:AoI_vs_ChnlAcs}
\end{figure}

In Fig.~\ref{fig:SimVerf}, we depict the network average AoI as a function of the packet arrival rate $\xi$, under different values of the deployment density $\lambda$.
From this figure, we first observe a close match between the simulation and analytical results, which verifies the accuracy of Theorem~\ref{thm:AoI_LCFS}.
Moreover, we note that the optimal update frequency that minimizes the average AoI is dependent on the particular value of $\lambda$.
Specifically, when $\lambda$ is small, the link pairs recede into the distance from each other and the packet transmissions can enjoy low level of interference because of the path loss.
This resembles a noise-limited scenario and, as pointed out by Remark~1, the average AoI can be reduced by increasing the update frequency at the source nodes.
On the contrary, when $\lambda$ becomes large, the network is densely deployed, in which the inter-link distances shrink and transmitters in geographical proximity can suffer from interference that results in transmission failures.
As such, with an increase of packet arrival rate, not only more link pairs are activated but, more crucially, additional failure packet deliveries and retransmissions are incurred, which prolongs the active period of the nodes. These together slow down the packet successful decoding process at each individual link and deteriorate the information freshness over the network.
In consequence, an optimal arrival rate exists that balances the tradeoff between the information freshness at the source nodes and the interference level across the network.
This observation shows an unconventional behavior of the AoI in random access networks employing LCFS queueing disciplines.

Fig.~\ref{fig:AoI_vs_ChnlAcs} plots the average network AoI for fixed $\lambda = 5 \times 10^{-2}$ as a function of the channel access probability $p$, under various packet arrival rates. We can see that in the situation of infrequent packet arrivals, the average AoI declines steadily as the channel access probability increases.
It is worth noting that this observation poses a dissent on the conclusions drawn from conflict graph models \cite{CheGatHas:19}, where the ALOHA protocol is asserted to be optimal for minimizing average AoI in the light traffic condition. The reason for such a difference is that under the SINR model, for small update frequency, the aggregated interference at each node is low and hence there is no necessity to reduce the channel use, which will, in turn, downgrade the packet successful decoding rate and deteriorate the AoI.
On the other hand, when the packet arrival rate is high, we can see that there exists an optimal channel access probability which minimizes the average AoI. This is because the defection of interference on the service rate is more devastating in this scenario, and adopting the ALOHA protocol is beneficial to striking a balance between information freshness at the transmitters and the overall interference level.
The figure also indicates that in order to achieve a small average AoI across the network, one should tune the update arrival frequency to a high level and adopt ALOHA to control the channel access.

\section{Conclusion}
In this work, we have developed a theoretical framework for the understanding of AoI performance in random access networks. We have used a general model that accounts for the channel gain and interference, dynamics of status updating, and spatial queueing interactions.
Our results have confirmed that the network topology has a direct and sweeping influence on the AoI.
Specifically, even when the transmitters employ an LCFS-R strategy for packet management, if the topology infrastructure is densely deployed then there exists an optimal rate of packet arrival that minimizes the average AoI. In addition, ALOHA is instrumental in further reducing the AoI, given the packet arrival rates are high.
However, when the network deployment density is low, the average AoI decreases monotonically with the packet arrival rate, and ALOHA cannot contribute to reducing the AoI in this scenario.

\begin{appendix}
\subsection{Proof of Lemma~\ref{lma:Cndt_AoI} } \label{apn:Cndt_AoI}
Let us consider a Geo/Geo/1 queueing system under the LCFS with preemption (LCFS-PR) discipline \cite{TriTalMod:19},
where the arrival and departure rates are set as $\xi$ and $p \mu^\Phi_0$, respectively. The AoI in this system evolves as follows:
\begin{align*}
\tilde{\Delta}_0(t \!+\! 1) =
\left\{
       \begin{array}{ll}
         \!\!  \tilde{\Delta}_0(t)  +   1, \qquad \qquad  ~~   \text{if transmission fails}, \\
         \!\!  \min\{ t  -  G_0(t), \tilde{\Delta}_0(t) \} + 1, \quad  ~~ \text{otherwise}
       \end{array}
\right.
\end{align*}
where $G_0(t)$ is the generation time of the packet delivered over the typical link at time $t$.
Then this system and the employed system in this paper possess the same AoI evolution statistics.

We denote by $M$ and $N$ the inter-arrival time and the total sojourn time in the queue, respectively, which are random variables.
As such, under the LCFS-PR discipline, the average AoI is given as \cite{TriTalMod:19}:
\begin{align} \label{equ:Gnl_AoI_L}
\mathbb{E}^0\big[ \bar{\Delta} \vert \Phi \big] = \frac{1}{2} \cdot \frac{ \mathbb{E}\big[ M^2 \big] }{ \mathbb{E}[ M ] } + \frac{\mathbb{E} \big[ \min(N, M) \big] }{ \mathbb{P}\big( N \leq M \big) } - \frac{1}{2}.
\end{align}
On the one hand, as $M \sim Geo(\xi)$ and $N \sim Geo( p \mu^\Phi_0)$, we have the following
\begin{align} \label{equ:Mmnt_M}
&\mathbb{E}[M] = \frac{1}{\xi}, \quad
\mathbb{E}[M^2] = \frac{ 2 - \xi }{ \xi^2 }, \\ \label{equ:Prob_NM}
& \mathbb{P}( N \leq M ) = 1 - \mathbb{E}\big[ ( 1 - p \mu^\Phi_0 )^M \big]
\nonumber\\
&\qquad \qquad ~~ = \frac{\mu^\Phi_0}{ 1 - ( 1 - p \mu^\Phi_0 ) ( 1 - \xi ) }.
\end{align}
On the other hand, since $M$ and $N$ are independent random variables, through simple calculations we have $\min(M,N) \sim Geo( 1 - (1-p\mu^\Phi_0)(1-\xi))$. Thus the following holds
\begin{align} \label{equ:Mmnt_min}
\mathbb{E}\big[ \min( M,N ) \big] = \frac{1}{ 1 - ( 1 - p \mu^\Phi_0 )(1-\xi) }.
\end{align}
The result in \eqref{equ:Cnd_AoI_LCFS} then follows from substituting \eqref{equ:Mmnt_M}, \eqref{equ:Prob_NM}, and \eqref{equ:Mmnt_min} into \eqref{equ:Gnl_AoI_L}.

\subsection{Proof of Lemma~\ref{lma:Cndt_ActProb} } \label{apn:Cndt_ActProb}
The evolution of the buffer state at a generic node $j$ can be modeled as a two-state Markov chain (empty/non-empty) with transition matrix given as follows:
\begin{align}
\mathbf{P} \!=\!
  \begin{bmatrix}
    ~1 - \xi &   \xi
    \nonumber\\
    ~ p \mu^\Phi_j (1-\xi) &  1 - p\mu^\Phi_j + p\mu^\Phi_j \xi
   \end{bmatrix}.
\end{align}
Let $\mathbf{v} = (v_0, v_1)$ denote the steady-state probability vector of the number of this Markov chain. Then, we have
\begin{align}
\mathbf{v}^{\mathrm{T}} = \mathbf{v}^{\mathrm{T}} \mathbf{P}, \\
v_0 + v_1 = 1.
\end{align}
Solving the above system of equations yields the following:
\begin{align}
v_0 = \frac{ p \mu^\Phi_j (1-\xi) }{ \xi + p \mu^\Phi_j (1-\xi) }, \\ \label{equ:buff_1}
v_1 = \frac{ \xi }{ \xi + p \mu^\Phi_j (1-\xi) }.
\end{align}
As such, the active state probability $a^\Phi_j$ can be obtained from \eqref{equ:buff_1} (the probability of having a non-empty buffer).

\subsection{Proof of Theorem~\ref{thm:Meta_SINR} } \label{apn:Meta_SINR}
For ease of exposition, let us denote $Y^\Phi_0 = \ln \mathbb{P}(\mathrm{SINR}_0 > \theta | \Phi)$. By leveraging Lemma~\ref{lma:Cnd_SucProb} and Lemma~\ref{lma:Cndt_ActProb}, we can calculate the moment generating function of $Y^\Phi_0$ as follows
\begin{align} \label{equ:MY_moment}
&\mathcal{M}_{ Y^\Phi_0 } (s) = \mathbb{E}\big[ ( \mu^\Phi_0 )^s \big]
\nonumber\\
& = e^{ - \frac{s \theta r^\alpha}{ \rho } } \mathbb{E}\Big[ \prod_{ j \neq 0 } \! \big( 1 - \frac{ a^\Phi_j }{ 1 \!+\! \Vert X_j \Vert^\alpha / \theta r^\alpha } \big)^s \Big]
\nonumber\\
& = e^{ - \frac{s \theta r^\alpha}{ \rho } } \mathbb{E}\Big[ \prod_{ j \neq 0 } \! \big( 1 - \frac{ 1 }{ 1 \!+\! \Vert X_j \Vert^\alpha / \theta r^\alpha } \cdot \frac{ p \xi }{ \xi \!+\! ( 1 \!-\! \xi ) p \mu^\Phi_j } \big)^s \Big]
\nonumber\\
& \stackrel{(a)}{=} e^{ - \frac{s \theta r^\alpha}{ \rho } } e^{- \lambda \int_{ \mathbb{R}^2 } \Big[ 1 - \big( 1 - \frac{1}{ 1 + \Vert x \Vert^\alpha / \theta r^\alpha } \cdot \frac{ p \xi }{ \xi + ( 1 - \xi ) p \mu_x } \big)^s \Big] dx }
\nonumber\\
& \stackrel{(b)}{=} \exp \! \bigg(\! - \frac{ s \theta r^\alpha }{ \rho } - \lambda \! \int_{ \mathbf{x} \in \mathbb{R}^2 } \sum_{ k = 1 }^{s} \! \binom{ s }{ k }   \frac{ (-1)^{k+1} d \mathbf{x} }{ ( 1 \!+\! \Vert \mathbf{x} \Vert^\alpha \!/ \theta r^\alpha )^k }
\nonumber\\
&\qquad \qquad \qquad \qquad \qquad \qquad \times \underbrace{\mathbb{E} \Big[ \big( \frac{ p \xi }{ \xi + ( 1 - \xi ) p \mu_{\mathbf{x}} } \big)^k \Big] }_{Q_1} \bigg),
\end{align}
where ($a$) follows by using the probability generating functional (PGFL) of PPP and ($b$) expands the expression via the binomial theorem.
A complete expression of \eqref{equ:MY_moment} requires us to compute the term $Q_1$, which however needs the CDF, $F(\cdot)$, of $\mu^\Phi_0$.
At this stage, let us assume the function $F(\cdot)$ is available. We can then expand the expectation term $Q_1$ and further reduce \eqref{equ:MY_moment} as we do below:
\begin{align} \label{equ:MY_reduce}
\mathcal{M}_{ Y^\Phi_0 }(s) &= \exp\!\bigg(\!\! - \frac{ s \theta r^\alpha }{ \rho } - \!\!\!  \int\limits_{ \mathbf{x} \in \mathbb{R}^2 } \sum_{ k = 1 }^{s} \! \binom{ s }{ k }   \frac{  (-1)^{k+1} \lambda d \mathbf{x} }{ ( 1 \!+\! \Vert \mathbf{x} \Vert^\alpha \! / \theta r^\alpha )^k }
\nonumber\\
&\qquad\qquad \qquad \qquad \times \int_{0}^{1}  \Big( \frac{ p \xi }{ \xi + ( 1 - \xi ) p t } \Big)^k F(dt) \bigg)
\nonumber\\
& = \exp\!\bigg(\!\! - \frac{ s \theta r^\alpha }{ \rho } - \lambda \pi r^2 \theta^\delta \sum_{ k = 1 }^{s} \! \binom{ s }{ k } \frac{ (-1)^{k+1} \! \int_{0}^{\infty} \! dv }{ ( 1 \!+\! v^{ \frac{\alpha}{2} } )^k }
\nonumber\\
&\qquad \qquad \qquad \qquad\qquad~\, \times \int_{0}^{1} \frac{ ( p \xi )^k F(dt) }{ \big[ \xi + ( 1 - \xi ) p t \big]^k }  \bigg).
\end{align}

Finally, by using the Gil-Pelaze theorem \cite{Gil}, we can derive the CDF of $\mu^\Phi_0$ as:
\begin{align}
F(u) &= \mathbb{P}( \mu^\Phi_0 < u ) = \mathbb{P}( Y^\Phi_0 < \ln u )
\nonumber\\
&= \frac{1}{2} - \frac{1}{\pi} \int_{0}^{\infty} \mathrm{Im} \big\{ u^{ - j \omega } \mathcal{M}_{Y^\Phi_0} ( j \omega ) \big\} \frac{d \omega}{ \omega }.
\end{align}
The statement readily follows by substituting \eqref{equ:MY_reduce} into the above equation.

\end{appendix}

\bibliographystyle{IEEEtran}
\bibliography{bib/StringDefinitions,bib/IEEEabrv,bib/howard_AoI_RAN}

\begin{thebibliography}{10}
\providecommand{\url}[1]{#1}
\csname url@samestyle\endcsname
\providecommand{\newblock}{\relax}
\providecommand{\bibinfo}[2]{#2}
\providecommand{\BIBentrySTDinterwordspacing}{\spaceskip=0pt\relax}
\providecommand{\BIBentryALTinterwordstretchfactor}{4}
\providecommand{\BIBentryALTinterwordspacing}{\spaceskip=\fontdimen2\font plus
\BIBentryALTinterwordstretchfactor\fontdimen3\font minus
  \fontdimen4\font\relax}
\providecommand{\BIBforeignlanguage}[2]{{%
\expandafter\ifx\csname l@#1\endcsname\relax
\typeout{** WARNING: IEEEtran.bst: No hyphenation pattern has been}%
\typeout{** loaded for the language `#1'. Using the pattern for}%
\typeout{** the default language instead.}%
\else
\language=\csname l@#1\endcsname
\fi
#2}}
\providecommand{\BIBdecl}{\relax}
\BIBdecl

\bibitem{KauYatGru:12}
S.~Kaul, R.~Yates, and M.~Gruteser, ``Real-time status: {How} often should one
  update?'' in \emph{Proc. IEEE INFOCOM}, Orlando, FL, Mar. 2012, pp.
  2731--2735.

\bibitem{KosPapAng:17}
A.~Kosta, N.~Pappas, and V.~Angelakis, ``Age of information: {A} new concept,
  metric, and tool,'' \emph{Foundations and Trends in Networking}, vol.~12,
  no.~3, pp. 162--259, 2017.

\bibitem{SunUysYat:17}
Y.~Sun, E.~Uysal-Biyikoglu, R.~D. Yates, C.~E. Koksal, and N.~B. Shroff,
  ``Update or wait: {How} to keep your data fresh?'' \emph{IEEE Trans. Inf.
  Theory}, vol.~63, no.~11, pp. 7492--7508, Nov. 2017.

\bibitem{YatKau:18}
R.~D. Yates and S.~K. Kaul, ``The age of information: {Real-time} status
  updating by multiple sources,'' \emph{IEEE Trans. Inf. Theory}, vol.~65,
  no.~3, pp. 1807--1827, Mar. 2019.

\bibitem{ZhaAraHua:19}
M.~Zhang, A.~Arafa, J.~Huang, and H.~V. Poor, ``How to price fresh data,'' in
  \emph{Proc. Modeling and Optimization in Mobile, Ad Hoc, and Wireless
  Networks (WiOpt)}, Avignon, France, Jun. 2019, pp. 1--8.

\bibitem{AraYanUlu:18}
A.~Arafa, J.~Yang, S.~Ulukus, and H.~V. Poor, ``Age-minimal transmission for
  energy harvesting sensors with finite batteries: Online policies,''
  \emph{IEEE Trans. Inf. Theory}, vol.~66, no.~1, pp. 534--556, Jan. 2020.

\bibitem{WuYanWu:18}
X.~Wu, J.~Yang, and J.~Wu, ``Optimal status update for age of information
  minimization with an energy harvesting source,'' \emph{IEEE Trans. Green
  Commun. Netw.}, vol.~2, no.~1, pp. 193--204, Mar. 2018.

\bibitem{BacSunUys:19}
B.~T. Bacinoglu, Y.~Sun, E.~Uysal, and V.~Mutlu, ``Optimal status updating with
  a finite-battery energy harvesting source,'' \emph{J. Commun. Netw.},
  vol.~21, no.~3, pp. 280--294, Jun. 2019.

\bibitem{AbdPapDhi:19}
M.~A. Abd-Elmagid, N.~Pappas, and H.~S. Dhillon, ``On the role of age of
  information in the internet of things,'' \emph{IEEE Commun. Mag.}, vol.~57,
  no.~12, pp. 72--77, Dec. 2019.

\bibitem{ZhoSaa:18}
B.~Zhou and W.~Saad, ``Optimal sampling and updating for minimizing age of
  information in the internet of things,'' in \emph{Proc. IEEE Global Commun.
  Conf. (Globecom)}, Abu Dhabi, United Arab Emirates, Dec. 2018, pp. 1--6.

\bibitem{XuYanWan:20}
C.~Xu, H.~H. Yang, X.~Wang, and T.~Q.~S. Quek, ``Optimizing information
  freshness in computing enabled {IoT} networks,'' \emph{{IEEE} Internet of
  Things Journal}, vol.~7, no.~2, pp. 971--985, Feb. 2020.

\bibitem{XuWanYan:20INFOCOM}
C.~Xu, X.~Wang, H.~H. Yang, H.~Sun, and T.~Q.~S. Quek, ``{AoI} and energy
  consumption oriented dynamic status updating in caching enabled {IoT}
  networks,'' in \emph{Proc. IEEE INFOCOM Workshop}, 2020.

\bibitem{HeYuaEph:16}
Q.~He, D.~Yuan, and A.~Ephremides, ``Optimizing freshness of information: {On}
  minimum age link scheduling in wireless systems,'' in \emph{Proc. Modeling
  and Optimization in Mobile, Ad Hoc, and Wireless Networks (WiOpt)}, Tempe,
  AZ, May 2016, pp. 1--8.

\bibitem{talak2018optimizing}
R.~Talak, S.~Karaman, and E.~Modiano, ``Optimizing information freshness in
  wireless networks under general interference constraints,'' \emph{arXiv
  preprint arXiv:1803.06467}, 2018.

\bibitem{CheGatHas:19}
X.~Chen, K.~Gatsis, H.~Hassani, and S.~S. Bidokhti, ``Age of information in
  random access channels,'' \emph{Available as ArXiv:1912.01473}, 2019.

\bibitem{CheGuLie:20}
H.~Chen, Y.~Gu, and S.-C. Liew, ``Age-of-information dependent random access
  for massive {IoT} networks,'' \emph{Available as ArXiv:2001.04780}, 2020.

\bibitem{HuZhoZha:18}
Y.~Hu, Y.~Zhong, and W.~Zhang, ``Age of information in {Poisson} networks,'' in
  \emph{Proc. Int. Conf. Wireless Commun. and Signal Process. (WCSP)},
  Hangzhou, China, Dec. 2018, pp. 1--6.

\bibitem{EmaElSBau:20}
M.~Emara, H.~ElSawy, and G.~Bauch, ``A spatiotemporal framework for information
  freshness in {IoT} uplink networks,'' \emph{{IEEE} Internet of Things
  Journal}, vol.~7, no.~8, pp. 6762--6777, Aug. 2020.

\bibitem{YanArafaQue:19}
H.~H. Yang, A.~Arafa, T.~Q.~S. Quek, and H.~V. Poor, ``Optimizing information
  freshness in wireless networks: A stochastic geometry approach,''
  \emph{{IEEE} Trans. Mobile Comput.}, 2020, Early Access.

\bibitem{BacBla:09}
F.~Baccelli and B.~Blaszczyszyn, \emph{Stochastic Geometry and Wireless
  Networks. Volumn I: Theory}.\hskip 1em plus 0.5em minus 0.4em\relax Now
  Publishers, 2009.

\bibitem{GupKum:00}
P.~Gupta and P.~R. Kumar, ``The capacity of wireless networks,'' \emph{IEEE
  Trans. Inf. Theory}, vol.~46, no.~2, pp. 388--404, Mar. 2000.

\bibitem{WanLinAdh:17}
Y.-P.~E. Wang, X.~Lin, A.~Adhikary, A.~Grovlen, Y.~Sui, Y.~Blankenship,
  J.~Bergman, and H.~S. Razaghi, ``A primer on {3GPP} narrowband internet of
  things,'' \emph{IEEE Commun. Mag.}, vol.~55, no.~3, pp. 117--123, Mar. 2017.

\bibitem{XioWuShe:17}
Y.~Xiong, N.~Wu, Y.~Shen, and M.~Z. Win, ``Cooperative network synchronization:
  Asymptotic analysis,'' \emph{IEEE Trans. Signal Process.}, vol.~66, no.~3,
  pp. 757--772, Feb. 2018.

\bibitem{TalKarMod:18}
R.~Talak, S.~Karaman, and E.~Modiano, ``Optimizing age of information in
  wireless networks with perfect channel state information,'' in \emph{Proc.
  Modeling and Optimization in Mobile, Ad Hoc, and Wireless Networks (WiOpt)},
  Shanghai, China, May 2018, pp. 1--8.

\bibitem{YanQue:19}
H.~H. Yang and T.~Q.~S. Quek, ``Spatiotemporal analysis for {SINR} coverage in
  small cell networks,'' \emph{IEEE Trans. Commun.}, vol.~67, no.~8, pp. 5520
  -- 5531, May 2019.

\bibitem{SanBac:17}
A.~Sankararaman and F.~Baccelli, ``Spatial birth--death wireless networks,''
  \emph{IEEE Trans. Inf. Theory}, vol.~63, no.~6, pp. 3964--3982, Jun. 2017.

\bibitem{ChiElSCon:19}
G.~Chisci, H.~ElSawy, A.~Conti, M.-S. Alouini, and M.~Z. Win, ``Uncoordinated
  massive wireless networks: Spatiotemporal models and multiaccess
  strategies,'' \emph{{IEEE/ACM} Trans. Networking}, vol.~27, no.~3, pp.
  918--931, Jun. 2019.

\bibitem{TriTalMod:19}
V.~Tripathi, R.~Talak, and E.~Modiano, ``Age of information for discrete time
  queues,'' \emph{Available as ArXiv:1901.10463}, 2019.

\bibitem{Gil}
J.~Gil-Pelaez, ``Note on the inversion theorem,'' \emph{Biometrika}, vol.~38,
  no. 3-4, pp. 481--482, Dec. 1951.

\end{thebibliography}

\end{document}